\documentclass[preprint,showpacs,preprintnumbers,amsmath,amssymb,superscriptaddress]{revtex4}

\usepackage{epsf}
\usepackage{graphicx}
\usepackage{sidecap}

\usepackage{color}

\def \beq {\begin{equation}}
\def \eeq {\end{equation}}
\begin{document}


\title{Electronic structure and relaxation dynamics in a superconducting topological material}


 




\author{Madhab~Neupane}
\affiliation {Condensed Matter and Magnet Science Group, Los Alamos National Laboratory, Los Alamos, NM 87545, USA}

\author{Yukiaki Ishida}
\affiliation {ISSP, University of Tokyo, Kashiwa, Chiba 277-8581, Japan}

\author{Raman Sankar} \affiliation{Center for Condensed Matter Sciences, National Taiwan University, Taipei 10617, Taiwan}

\author{Jian-Xin~Zhu}
\affiliation {Theoretical Division, Los Alamos National Laboratory, Los Alamos, NM 87545, USA}

\author{Daniel S.~Sanchez}
\affiliation {Joseph Henry Laboratory and Department of Physics, Princeton University, Princeton, New Jersey 08544, USA}

\author{Ilya~Belopolski}
\affiliation {Joseph Henry Laboratory and Department of Physics, Princeton University, Princeton, New Jersey 08544, USA}

\author{Su-Yang Xu}
\affiliation {Joseph Henry Laboratory and Department of Physics, Princeton University, Princeton, New Jersey 08544, USA}

\author{Nasser~Alidoust}\affiliation {Joseph Henry Laboratory and Department of Physics, Princeton University, Princeton, New Jersey 08544, USA}











\author{Shik~Shin}
\affiliation {ISSP, University of Tokyo, Kashiwa, Chiba 277-8581, Japan}

\author{Fangcheng Chou} \affiliation{Center for Condensed Matter Sciences, National Taiwan University, Taipei 10617, Taiwan}

\author{M.~Zahid~Hasan}
\affiliation {Joseph Henry Laboratory and Department of Physics,
Princeton University, Princeton, New Jersey 08544, USA}

\author{Tomasz~Durakiewicz}
\affiliation {Condensed Matter and Magnet Science Group, Los Alamos National Laboratory, Los Alamos, NM 87545, USA}

\date{18 June, 2013}
\begin{abstract}

{Topological superconductors host new states of quantum matter which show a pairing gap in the bulk and gapless surface states providing a platform to realize Majorana fermions. Recently, alkaline-earth metal Sr intercalated Bi$_2$Se$_3$ has been reported to show superconductivity with a T$_c$ $\sim$ 3 K and a large shielding fraction.  
Here we report systematic normal state electronic structure studies of Sr$_{0.06}$Bi$_2$Se$_3$ (T$_c$ $\sim$ 2.5 K) by performing photoemission spectroscopy. Using angle-resolved photoemission spectroscopy (ARPES), we observe a quantum well confined two-dimensional (2D) state coexisting with a topological surface state in Sr$_{0.06}$Bi$_2$Se$_3$. Furthermore, our time-resolved ARPES reveals 
the relaxation dynamics showing different decay mechanism between the excited topological surface states and the two-dimensional states. Our experimental observation is understood by considering the intra-band scattering for topological surface states and an additional electron phonon scattering for the 2D states, which is responsible for the superconductivity.  Our first-principles calculations agree with the more effective scattering and a shorter lifetime of the 2D states. Our results will be helpful in understanding low temperature superconducting states of these topological materials.}





\end{abstract}
\date{\today}
\maketitle



The discovery of three-dimensional (3D) topological insulators (TIs) in bismuth-based semiconductors
has attracted considerable amount of research interest in condensed matter physics \cite{Moore, Hasan, SCZhang, Hsieh, Neupane_1, Hasan_review_2, Hasan_hetro, Neupane_2}. In these materials, the bulk has a full energy gap whereas the surface possesses an odd number of Dirac-cone electronic states,
where the spin of the surface electrons is locked to their linear momentum \cite{Hasan, SCZhang}. 
Such unique properties make these materials alluring not only for studying various fundamental phenomena in condensed matter physics and particle physics, but also provide high potential for applications ranging from spintronics to quantum computation \cite{Hasan, SCZhang,Hasan_hetro}.
More specifically, it has been recently predicted that the long-sought-out Majorana fermions can be realized  in the  interface between a topological insulator and a superconductor \cite{Liang_SC_1, Hasan_hetro, Liang_SC, Vishwanath_PRL2011}. There has been a great effort in condensed matter physics to realize the Majorana fermion quasiparticle states associated with topological superconductivity.

Although the experimental realizations of topological superconductors in real materials have remained
considerably limited, the search for the bulk topological superconductors is an ongoing project in condensed-matter physics. To date, possible topological superconductivity has been suggested in Cu-intercalated
Bi$_2$Se$_3$ (Cu$_x$Bi$_2$Se$_3$) \cite{Cava_CuBi2Se3, Ando_CuBi2Se3,Wray_CuBi2Se3}, highly-pressurized Bi$_2$Te$_3$ and Sb$_2$Te$_3$ \cite{press_Bi2Te3}, and Bi$_2$X$_3$ (X= Se, Te) thin films grown on superconducting substrates \cite{Hasan_hetro, Bi2Te3_hetro}. After the discovery of superconductivity in Cu$_x$Bi$_2$Se$_3$, an enormous amount of research effort was devoted to realize the  first bulk topological superconductor \cite{Ando_CuBi2Se3_TI, STM_CuBi2Se3}. However, a clear signature of topological superconductivity in this material remains still elusive mainly due to the relatively low superconducting volume fraction of the sample. Recently, it was reported that the intercalation of an alkaline-earth metal Sr in the well-studied topological insulator
Bi$_2$Se$_3$ (Sr$_x$Bi$_2$Se$_3$) shows a superconducting state with T$_c$ $\sim$ 3.0 K and a large superconducting volume fraction ($\sim$ 90\%), providing a more ideal platform to realize a topological superconductor \cite{Shruti_SrBi2Se3, Liu_SrBi2Se3}. However, a detailed electronic structure of this new compound has not been reported. 
A detailed systematic high-resolution angle-resolved photoemission spectroscopy (ARPES) study is needed to investigate the nature of the surface states in the normal state of Sr intercalated Bi$_2$Se$_3$.
Such a characterization is a necessary first step towards understanding the relationship between superconductivity and the topological properties of this new compound.
Furthermore, the occurrence of superconductivity (SC) in topological insulators is a subject of strong current interest, because of the theoretical predictions for the possible observation of exotic and unexplored excitations. ARPES and time-resolved ARPES (TRARPES) have provided detailed pictures of the topological surface states, and appear as the most appropriate tools to reveal the unusual features of topological superconductivity.

Here, we report the normal ARPES and time-resolved ARPES investigation of the detailed electronic structure of Sr intercalated Bi$_2$Se$_3$. Using normal ARPES, we observe the coexistence of the two-dimensional (2D) quantum well states and single Dirac cone topological surface states present in this system with the Dirac point located around 450 meV below the Fermi level. More importantly, by using time-resolved ARPES (TRARPES), we observe different relaxation mechanisms for the 2D state and topological surface state. The observed effect can be understood by considering an additional scattering term for 2D states.  Our theoretical calculations show the more effective scattering and a shorter lifetime of the 2D states, which is consistence with the experimental observation.
Our results provide critical knowledge necessary for realizing and understanding the topological superconductivity in this system which helps to demonstrate the Majorana fermion associated with topological superconductivity.


\bigskip
\bigskip
\textbf{Results}
\newline

\textbf{Transport characterization of Sr$_{0.06}$Bi$_2$Se$_3$}

We start our discussion by presenting transport characterizations of the samples used in our spectroscopy measurements.  Fig. 1a shows the temperature dependent in-plane resistivity of Sr$_{0.06}$Bi$_2$Se$_3$.
The onset of superconducting transition is about T$_c \sim$ 2.5 K. The observation of a narrow superconducting transition width  suggests that the samples used in our measurements are of high quality (see the inset of Fig. 1a for picture of a sample used for measurements).  Fig. 1b shows the temperature dependent magnetic susceptibility measured with a magnetic field parallel to the in-plane of the sample. The shielding volume fraction at 0.5 K is estimated to be about 90$\%$. 
The high superconducting volume fraction of these samples provides a new and better opportunity to study topological superconductivity. Furthermore, results obtained from transport characterization are consistent with recent reports \cite{Liu_SrBi2Se3, Shruti_SrBi2Se3}.


\bigskip
\textbf{Observation of 2D states in Sr$_{0.06}$Bi$_2$Se$_3$}

Fig. 2a shows the ARPES measured dispersion map of Sr$_{0.06}$Bi$_2$Se$_3$  using normal ARPES setup with photon energies of 22 eV and 24 eV. A sharp V-shaped topological surface states is observed, where the Dirac point is located about 450 meV below the Fermi level. Most importantly, we observe the coexistence of the topological surface states and a 2D states on the surface of the Sr$_{0.06}$Bi$_2$Se$_3$ \cite{David_Nat09, Hofmann_1, Hofmann_2}. These 2D states originate from the bulk band-bending near the surface due to the Sr-intercalation.  The left panel of Fig. 2b shows the ARPES spectra measured in a pump probe setup using a 6 eV photon source in the absence of a pump pulse. Using this experimental setup, the coexisting topological surface states and 2D states are also observed. 
From these two sets of experimental results, it is clear that the presence of the 2D quantum well state is a generic property of the Sr intercalated prototypical topological insulator Bi$_2$Se$_3$.  


\bigskip

\textbf{Carrier dynamics of Sr$_{0.06}$Bi$_2$Se$_3$}

Now we turn our focus onto the TRARPES results. We note that our results constitute the first pump-probe ARPES 
in a system with coexisting topological surface states and 2D states.
Figs. 2b(right panel) and Fig. 3a-b show TRARPES spectra of the Sr$_{0.06}$Bi$_2$Se$_3$  sample at the representative delay time ($d.t.$) values before and after the pump pulse.
Without a pump pulse, the chemical potential of the Sr$_{0.06}$Bi$_2$Se$_3$ sample cuts the topological surface states and 2D quantum well states, which shows its bulk $n$-type metallic character.
In order to systematically study how the excited electronic states relax, we present the transient ARPES spectra as a function of $d.t.$ in Fig. 3c. For a metallic sample of Sr$_{0.06}$Bi$_2$Se$_3$, the population of the excited topological surface states and 2D quantum well state relaxes within $5$ ps.  It is consistent with the short optical life-time of a few picoseconds ($10^{-11}$-$10^{-12}$ s) for the Dirac surface states reported previously for bulk metallic samples  \cite{Sobota_1, Hajlaoui, Wang, Crepaldi_1, Mi, Marsi, Neupane_TRARPES}.  The short life-time is expected because
these experiments were performed with bulk metallic TI samples, which means both the surface states and the bulk conduction bands are present at the chemical potential. 
We note that the spread of the intensity into the unoccupied side is less pronounced than those observed
in less metallic TIs. In the latter case, the spread is visible as far as 1 eV above $E_F$.
Furthermore, the rise of intensity is instantaneous, or resolution limited, and is strongly contrasted to
the cases of less metallic TIs. In the latter case, topological surface states are indirectly populated and shows delayed filling of $\gtrsim$0.5 ps.


Most importantly, we observe a different decay mechanism for the excited topological surface states than the 2D state (see Fig. 3c-d).  If the scattering of the carriers in both the topological surface states and 2D states  was strong  and of similar nature, then the decay profile of the two should overlap. 
In order to get an insight into the decay mechanism, we use simple single and double exponential decay fitting functions (see Fig. 3d). When fitted with the double exponential decay function, both Dirac quasiparticle populations L-{SS} and R-{SS}, integrated as shown in Fig. 3b, produce almost identical results, with both components of the exponential fitting giving the same amplitude and decay constants for both components, and only a slightly difference between L and R parts, as shown in Table below. We conclude that only one decay constant is present in Dirac quasiparticle band and we use a single exponential decay function here. However, the 2D surface state does not converge well when fitted with one decay channel. It shows a short-time decay channel followed by almost the same, longer decay as in Dirac bands, and hence requires a two-component exponential fit.
The values obtained from the fitting (see Fig. 3d) are shown in the following Table:


\begin{center}
 \begin{tabular}{||c c c c||} 
 \hline
 Band & Decay constants (ps) & r$^2$ & Amplitude (A)\\ [0.5ex] 
 \hline\hline
 L-{SS} & $\tau_1 =\tau_2$=1.58 & 0.99 & A=1.382\\ 
 \hline
 R-{SS} & $\tau_1 =\tau_2$ =1.69 & 0.99 & A=1.347 \\
 \hline
 2D state & $\tau_1$=0.70 and $\tau_2$=1.46 & 0.99 & A1=0.33, A2=1.14 \\
 \hline
\end{tabular}
\end{center}

The performed fitting indicates that within the Dirac bands we are looking at a simple, one-component scattering mechanism, while in the 2D surface state part, there are two channels. One of those channels is fast, not present in the Dirac bands, and 
the second one is slow and similar to the mechanism found in the Dirac bands.
The latter indicates that there is a weak inter-band coupling that synchronizes the decay at $\gtrsim$1 ps between the Dirac and 2D bands.



\bigskip
\bigskip
\textbf{Discussion}
\newline

To explain the observed difference in life-time, we first ascribe the picosecond decay time to the coupling of surface Dirac quasiparticles and the Sr-doping induced 2D state to the surface phonons. In this time regime, the scattering of intra-band quasiparticles off the phonon modes is the dominant contribution to quasiparticle lifetime. Theoretically,  the decay time is then written as follows:

$1/\tau = \pi \sum_{\mu} \vert g_{\mu} \vert^2 [f_{BE}(\Omega_\mu) + f_{FD}(\Omega_\mu)][N(\Omega_\mu) + N(-\Omega_\mu)]$, where $\Omega_\mu$ is the phonon mode from a branch $\mu$, $g_{\mu}$ is the coupling strength between this mode and quasiparticles,
$f_{BE}(E) = 1/(e^{E/k_{B}T}-1)$ and $f_{FD}(E)=1/(1+e^{E/k_{B}T})$ are the Bose-Einstein and Fermi-Dirac distribution function, $N(E)$ is the quasiparticle density of states. Based on our photoemission spectroscopy measurement, a schematic drawing of surface density of states for the Dirac quasiparticles and the 2D states are shown in Fig. 3e.  The above equation shows clearly that for a given phonon mode energy, the decay time is inversely proportional to the quasiparticle density of state (DOS) at this energy $\Omega_\mu$ and its imaging $-\Omega_\mu$.
Experiments on pure Bi$_2$Se$_3$ \cite{Zhu} have revealed that the surface phonon modes mostly strongly coupled to the surface Dirac quasiparticles are located at near 7.4 meV, the location of which and its imaging are marked with dashed lines in Fig. 3e. Our measurements indicate that the 2D surface states are located between the Fermi level and roughly 200 meV below E$_{F}$. Based on these observations we propose that the phonon scattering channel which can be linked to the intra-band scattering is the same for all bands, while the 2D states are open to an additional scattering mechanism, possibly related to local vibration modes around the Sr - dopants. 
Theoretically, we have performed first-principles electronic structure calculations to the end member SrBi$_2$Se$_3$ in the R$\bar{3}$m structure. Although this structure phase has much higher concentration of Sr, which may affect the electronic states near the Fermi energy, it should give a reasonable approximation for the local vibrational properties of  Sr dopant along the $c$-axis. The energy of this vibrational mode of Sr is found to be 14.12 meV and is schematically marked with the thick magenta lines in Fig. 3e. The coupling of this mode to the 2D electronic states, through the hybridization of the 6p$_z$ orbital of Bi ion with the 5s orbitals of Sr ion,  leads to more effective scattering and a shorter lifetime of the 2D states.


In order to reveal the mechanism of superconductivity, we consider the electron-phonon coupling ($\lambda$) in Bi$_2$Se$_3$ determined from ARPES experiments ranging from 0.08 \cite{Valla_PRL2012} to 0.25 \cite{Hofmann_PRB2012}. ARPES approach represents the electron aspect of the electron-phonon coupling, incorporating 
the quasiparticle renormalization effect contributed from all relevant phonon modes in one measured slope of temperature dependence.
The inelastic helium-atom surface scattering measurements \cite{HAS_PRL2012} was recently used to look at the coupling purely from the phonon perspective for a surface-phonon branch, and the value found was 0.43, greater than those estimated by ARPES. The value of 0.43 represents a lower bound on the value of the actual coupling and is used here. Using the McMillan formula \cite{Mcmillan_PRB1968} of the Migdal-Eliashberg theory \cite{Eliashberg_1960}, a value of 0.1 for Coulomb pseudopotential and a value of 185K for Debye temperature \cite{Debye_2012} we obtain T$_c$=1K for $\lambda$ = 0.43, and T$_c$=2.5K for $\lambda$ = 0.54, in agreement with experimental T$_c$ of around 2.5K and the estimate of $\lambda$ with inelastic helium-atom surface scattering measurements \cite{HAS_PRL2012}. For completeness, we also calculate the simple BCS value, which  does not include the retardation effect of the actual electron-phonon interactions as described by the Migdal-Eliashberg theory. The BCS value is 9K for the same set of parameters, which still within a factor of 4 from the experimental value of T$_c$.
 The result obtained from Migdal-Eliashberg theory is in better-than-expected agreement with experimental value of T$_c$. We interpret this finding as evidence for phonon-mediated mechanism of superconductivity in Sr-doped Bi$_2$Se$_3$.



Sr intercalated Bi$_2$Se$_3$ system serves a platform to study the interconnection between topology and superconductivity.
It provides a natural interface between a spin-polarized topological surface state and superconductivity. Majorana fermions are predicted to occur at the surface of a topological insulator in proximity of superconductivity \cite{Liang_SC_1}, and may also occur at such a natural interface.
The strong spin splitting of the TI allows for p-wave Cooper pairing on
the surface of this compound in spite of the s-wave superconductivity in the
bulk. We note that because the T$_c$ of Sr$_{0.06}$Bi$_2$Se$_3$ is large enough to study the zero-biased peak (ZBP) by
scanning tunneling microscopy (STM), this compound is a promising candidate
to get insights of the interaction between superconductivity and
topological surface states.

In conclusion, by using ARPES and TRARPES we study the normal state properties of the Sr intercalated Bi$_2$Se$_3$. We find the coexistence of the topological surface states and 2D quantum well states at the surface of  Sr$_{0.06}$Bi$_2$Se$_3$. Furthermore, using TRARPES, we find the different decay mechanisms for the excited TI surface states and 2D quantum wells, which can be understood by considering the different scattering mechanisms. 
Our first-principles calculations show the more effective scattering leading to a shorter lifetime of the 2D states, which is consistence with the experimental observation.
Our systematic study will be helpful in understanding the topological superconducting properties of this material, which helps to realize
the properties of Majorana fermion quasiparticle states associated with topological superconductivity.



\bigskip
\bigskip
\textbf{Methods}
\newline

\textbf{Sample growth and characterization}

Single crystalline samples of Sr-intercalated Bi$_2$Se$_3$ topological insulator (Sr$_{0.06}$Bi$_2$Se$_3$) used in our measurements were grown using the Bridgman method and characterized by transport methods, which is detailed elsewhere \cite{Cava_CuBi2Se3, Shruti_SrBi2Se3, Liu_SrBi2Se3}.

\textbf{Spectroscopic measurements} 

The TRARPES setup at the Institute for Solid State Physics (ISSP) in the University of Tokyo consisted of an amplified Ti:sapphire laser system delivering h$\nu$= 1.47 eV pulses of 170-fs duration with 250-kHz repetition and a hemispherical analyzer \cite{Ishida}. A portion of the laser was converted into h$\nu$= 5.9 eV probing pulses using two non-linear crystals, $\beta-$BaB$_2$O$_4$ (BBO), and the time delay ($d.t.$) from the pump was controlled by a delay stage. 
The diameters of the pump and probe beams were 0.5 and 0.3 mm, respectively, and the time resolution of the pump-and-probe measurement was 250 fs.
The pump and probe pulses were $s$ and $p$-polarized, respectively. The probe intensity was lowered so that the space-charge induced shift of the spectrum is less than 5 meV. The TrARPES spectra were recorded with the energy resolution of about 15 meV. Multi-photon photoelectrons due to the
pump pulse were not observed in the dataset presented herein.
The base pressure of the photoemission chamber was 5$\times$10$^{-11}$ Torr. 
With this TrARPES setup, our Sr$_{0.06}$Bi$_2$Se$_3$ samples were cleaved and measured at low temperature ($\sim$ 8 K).

Normal ARPES measurements were performed at the SIS-HRPES end-station of the Swiss Light Source with a Scienta R4000 hemispherical electron analyzer. The minimum sample temperature was 17 K and the pressure during measurements was better than 5$\times$10$^{-11}$ mbar. The energy and angular resolution were set to be better than 20 meV and 0.2$^{\circ}$ for the measurements with the synchrotron beamline.

\textbf{Electronic structure calculations}

Our calculations are carried out within the density functional theory as implemented in the {\em  ab initio} package VASP \cite{Kress_1}. The Perdew-Burke-Ernzerhof exchange-correlation functional \cite{Perdew} in the projector augmented plane-wave potential format \cite{Blochl, Blochl_1} is used and the spin-orbit coupling is taken into account. The first Brillouin zone is sampled with 8$\times$ 8$\times$1 k-points using the Monkhorst-Pack grid.
The vibration analysis of Sr atoms is done after a full structure relaxation with the converged lattice parameters a=b=4.535 \AA  \hspace{0.0cm} and c=31.38 \AA. By taking into the concentration difference of Sr, these lattice constants agree reasonably well with those Sr$_x$Bi$_2$Se$_3$ in the dilute limit \cite{Liu_SrBi2Se3, Shruti_SrBi2Se3}.


\bigskip
\bigskip
\bigskip
\hspace{0.5cm}
\textbf{Acknowledgements}

M.N. is supported by LANL LDRD Program. T.D. is supported by Department of Energy, Office of Basic Energy Sciences, Division of Materials Sciences, and by NSF IR/D program.
The work at Princeton and synchrotron X-ray-based measurements are supported by the Office of Basic Energy Sciences, US Department of Energy (grants DE-FG-02-05ER46200, AC03-76SF00098 and DE-FG02-07ER46352).
S.S. and Y.I. at ISSP in University of Tokyo acknowledge support from KAKENHI Grants number
23740256 and 2474021. J.-X. Zhu is supported by the DOE Office of Basic Energy Sciences.
We thank Plumb Nicholas Clark for beamline assistance at the SLS, PSI. 

\*Correspondence and requests for materials should be addressed to M.N. (Email: mneupane@lanl.gov).

\newpage

\begin{figure}
\centering
\includegraphics[width=14.50cm]{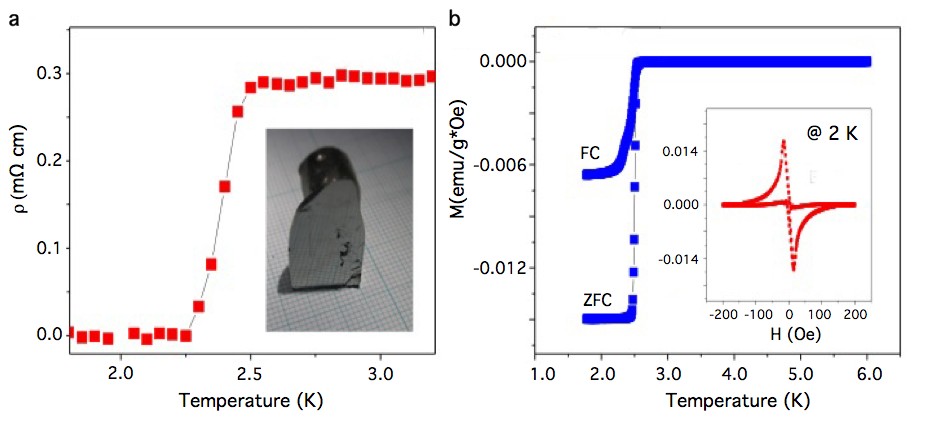}
\caption{\textbf{Transport characterizations of Sr$_{0.06}$Bi$_2$Se$_3$}. 
\textbf{a,} Resistivity vs temperature of Sr$_{0.06}$Bi$_2$Se$_3$. Inset shows the picture of a sample used for measurements. \textbf{b,} Temperature dependence of magnetic susceptibility for the Sr$_{0.06}$Bi$_2$Se$_3$ sample measured with applied magnetic field (10 Oe) parallel to the in-plane of the sample. The shielding volume fraction estimated from the zero-field cooling (ZFC) process is about 90$\%$. The inset shows the magnetic field dependence of magnetization measured at 2.0 K.}
\end{figure}

\newpage

\begin{figure*}
\centering
\includegraphics[width=18.00cm]{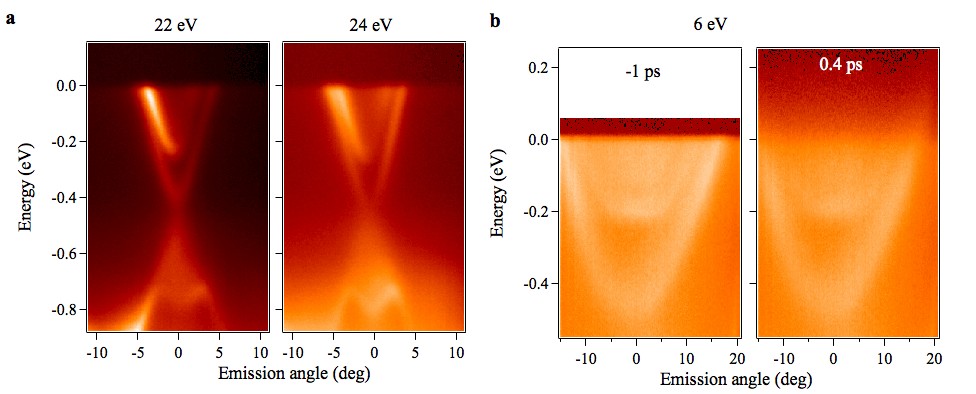}
\caption{\textbf{Observation of 2D states in Sr$_{0.06}$Bi$_2$Se$_3$}. \textbf{a,} Dispersion map of Sr$_{0.06}$Bi$_2$Se$_3$  along the $\bar{\textrm{K}}$-$\bar\Gamma$-$\bar{\textrm{K}}$ high-symmetry direction  recored by using incident photon energy of 22 eV and 24 eV at a temperature of 17 K. This dataset was obtained with normal ARPES setup. \textbf{b,} Dispersion maps measured in TRARPES setup with 6 eV laser source. Left panel shows the ARPES band dispersion of Sr$_{0.06}$Bi$_2$Se$_3$ at a negative time delay along the $\bar{\textrm{K}}$-$\bar\Gamma$-$\bar{\textrm{K}}$ high-symmetry direction. Right panel shows the dispersion maps for positive time delay. At both experimental setup, the 2D quantum well states are observed.  } 
\end{figure*}

\begin{figure*}
\centering
\includegraphics[width=14.00cm]{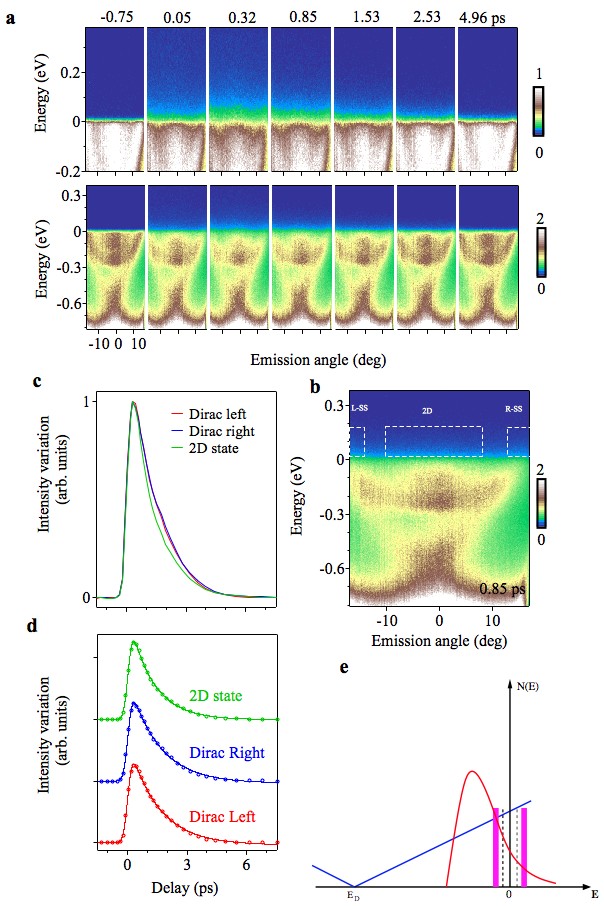}
\caption{\textbf{Relaxation mechanism.} }
\end{figure*}

\addtocounter{figure}{-1}
\begin{figure*}[t!]
\caption{\textbf{a,} TRARPES images of Sr$_{0.06}$Bi$_2$Se$_3$ before and after the pump pulse. 
 Top panels show band dispersions obtained with the difference to the image before pumped, and bottom panels show the time-evolution sprectra. \textbf{b,} ARPES band dispersion of Sr$_{0.06}$Bi$_2$Se$_3$ at a positive time delay along the $\bar{\textrm{K}}$-$\bar\Gamma$-$\bar{\textrm{K}}$ high-symmetry direction. The blue rectangles represent the integration window of transient photoemission intensity for Dirac surface state and 2D states.
\textbf{c,} Ultrafast evolution of the population of 2D quantum well states (black curve) and surface states (red for left surface state and blue for right surface state) for Sr$_{0.06}$Bi$_2$Se$_3$. 2D quantum-well states relax faster than topological surface states. \textbf{d,} Exponential fitting for the decay curves; fitting parameters are given in the Table. We only consider the decay of the carriers. The rise of the carriers are almost identical within the time-resolution.
\textbf{e,} Schematic view of the decay mechanism, where blue line represents the density of Dirac quasiparticles and red line corresponds to the density of surface 2D state, while dashed grey lines indicate the energy of the phonon mode. The surface phonon modes mostly strongly coupled to the surface Dirac quasiparticles are located at near 7.4 meV, marked with dashed lines. The thick magenta lines represent the vibrational mode of Sr, which is found to be about 14.12 meV.}
\end{figure*}

\end{document}